\documentclass[aps,prfluids,onecolumn,preprint,superscriptaddress]{revtex4-2}

\usepackage{graphicx}   
\usepackage{dcolumn}    
\usepackage{bm}         
\usepackage{amsmath}
\usepackage{amssymb}
\usepackage[usenames]{color}      

\begin{document}
	
\title{Flow structures and vertical transport in tilting salt finger with a background shear}
	
\author{Junyi Li}
\affiliation{State Key Laboratory for Turbulence and Complex Systems and Department of Mechanics and Engineering Science, Beijing Innovation Center for Engineering Science and Advanced Technology, College of Engineering, Peking University, Beijing 100871, China}
	
\author{Yantao Yang}\email{yantao.yang@pku.edu.cn}
\affiliation{State Key Laboratory for Turbulence and Complex Systems and Department of Mechanics and Engineering Science, Beijing Innovation Center for Engineering Science and Advanced Technology, College of Engineering, Peking University, Beijing 100871, China}

\date{\today}
	
\begin{abstract}
In this work we study the fingering double diffusive convection, namely the buoyancy-driven convection flow within a fluid layer experiencing an unstable salinity gradient and a stable thermal gradient. Especially, we investigate the influences from a background shear with uniform strength. Linear stability analysis indicates that the unstable modes shift from a circular shape to a sheet-like shape as the shear becomes stronger. Three dimensional direction numerical simulations are conducted for five groups of cases, each of which has the same combination of thermal and salinity gradients (measured by corresponding Rayleigh numbers), and gradually increasing shear strength. Simulation results reveal that a very weak shear organizes the salt fingers into a very regular pattern, which can enhance the salinity flux. This enhancement effect, however, reduces as the Rayleigh number increases. As the shear further strengthens, the dominant structures change from salt fingers to salt sheets, and the coherence length scale increases in the streamwise direction. Meanwhile, salinity and heat fluxes decrease. These results suggest that even a weak shear can notably alter the morphology and transport properties of fingering double diffusive convection.
\end{abstract}
	
\maketitle
	
\section{Introduction}\label{sec:intro}

When fluid density depends on two scalar components, buoyancy-driven convection happens in the form of double diffusive convection (DDC). DDC is ubiquitous in the Ocean since the density of seawater is determined mainly by temperature and salinity \citep{turner1974, schmitt1994, you2002}. In the upper water of many tropical and subtropical oceans, salinity and temperature decreases as the depth increases \citep{schmitt2005}. In these regions DDC happens in the finger regime, since the slow diffusing salinity gradient drives the flow and fast-diffusing temperature gradient stabilizes the flow \citep{stern1960}. DDC is very important for the vertical mixing and transport, and an excellent summary of the field can be found in the recent book of Radko~\citep{radko2013}.

In real ocean environments, horizontal currents are omnipresent and inevitably interact with DDC motions, especially for the finger regime. For instance, observations have found the nearly horizontal small-scale laminae in the salt-finger unstable region of the western North Atlantic~\citep{kunze1987optical, kunze1990evolution}. The authors assume these structures to be the salt fingers titled by background shear. Investigations also confirm that the moderate shear indeed appears in the high-gradient interfaces of such regions, with typical Richardson number of $Ri\sim6$~\citep{gregg1987}. Other evidences of the existence of tilted fingers include the vertical and sloping filaments found in the eastern North Atlantic~\citep{stlaurent1999}. All these observations indicate that the morphology and transport properties of fingers will be affected by the shear in the Ocean. 

Over the years people have confirmed the abundant interactions between the salt fingers and the background shear. The first systematic study, including experiments and theoretical analyses, of the shear effects on salt fingers is conducted by~\citet{linden1974}. The author states that the steady uniform shear will dampen the DDC disturbances along the streamwise direction, leading to the two-dimensional structure, namely, the salt sheet. Through the linear stability analysis, such two-dimensional rolls are proved to be the most unstable mode in a finger regime under the shear with no inflection points~\citep{thangam1984}. For the inflectional shear, the flow is susceptible to the Kelvin–Helmholtz (KH) instability, which can coexist with the salt sheets~\citep{smyth2007instability}. The fully nonlinear numerical simulations show that the KH instability has a more lenient criteria with $Ri$ exceeding unity under the DDC motions~\citep{radko2011}. Other three-dimensional direct numerical simulations (DNS) also demonstrate that two secondary instabilities will appear after the formation of salt sheets, which are called zig-zag and tip modes~\citep{kimura2007direct,smyth2011mixing}. The former may explain the origin of the oceanic horizontal bands mentioned above~\citep{kunze1987optical,smyth2011c}. As the morphology of salt fingers undergoes various variations under the action of background shear, their transport efficiency also changes accordingly.

The key question about the sheared salt fingers is how the scalar transport capacity varies with the shear strength. It is natural to assume that the heat and salinity fluxes will decease to a two-dimensional level when the salt sheets form, and \citet{radko2015} reach this conclusion by DNS with the stochastic shear related to the internal wave in the real ocean. However, this is not always the case when the salt sheets do not totally appear, or the secondary instabilities occur. In addition, the experimental and numerical settings will also affect the transport properties a lot. For instance, the experiments of \citet{linden1974} reveal that the fluxes are even enhanced under the shear, which may due to the working fluid is flowing in and out of the domain constantly. \citet{fernandes2010} conduct experiments and observe that the salinity flux decreases with the shear strength, and the heat flux eventually decreases to a total molecular diffusive mode. Through DNS in a sharp high-gradient interface, Kimura and Smyth analyze the transport properties for the salt sheets in which secondary instabilities appear, and they conclude that the effective diffusivities of heat and salt decrease with the shear strength and the density ratio~\citep{kimura2007direct,kimura2011}. Recently~\citet{sichani2020} conduct a series of DNS for the bounded sheared fingers within two horizontal plates, and they still find the reduced salinity fluxes. Some models are proposed to explain the declined fluxes~\citep{kunze1990evolution,wells2001}, but a general theory is still lack for the various morphology of the sheared fingers. Moreover, the Richardson number in these studies are constrained in relatively small values. The situation with weak shear (or large Richardson number) is rarely explored.

It may be natural to assume that weak shear should not have notable influence. However, in this study we will show that even a relatively weak shear can significantly alter the flow morphology and fluxes. Based on our previous work for fingering DDC~\citep{ddcjfm2016}, here we investigate the fingering DDC under the influence of background shear flow with a series of large Richardson numbers of $Ri \sim 10-10^6$. Different density ratios and Rayleigh numbers are simulated. The lower bound of $Ri$ is for the stage that the salt sheets just begin to appear, and is far from the criteria for KH instability. Thus in the current study we do not consider the secondary instabilities or the KH instability. We focus on the changes of the finger structures and the transport fluxes in the vertical direction under the relatively weak shear. 

This paper is organized as follow. In section~\ref{sec:eqn} we introduce the governing equations and the related control domain. Next we conduct the linear stability analysis in section~\ref{sec:stablity}. Then section~\ref{sec:dns} presents the typical DNS results. Finally we give the conclusions in section~\ref{sec:con}.

\section{Governing equations}\label{sec:eqn}

We consider the incompressible Navier-Stokes equations for a fluid layer bounded by two horizontal plates which are separated by a height of $H$. Let $z^*$ be the normal direction of the plates ($z^*=0$ at the bottom and $z^*=H$ at the top), while $x^*$ and $y^*$ are the horizontal directions. Hereafter, the asterisk $*$ stands for the quantities in dimensional form,  and the $(x^*,y^*,z^*)$ directions are referred as spanwise, streamwise and vertical directions, respectively. Gravity is oriented along the negative vertical direction. An uniform and steady background shear is sustained in the streamwise direction as $\mathbf{U}_s=S(z^*-H/2)\textbf{{e}}_y$. Here $S$ is the shear strength and $\mathbf{e}_y$ is the streamwise unit vector, respectively. The total velocity then reads $\textbf{u}^*=\hat{{\textbf{u}}}+\textbf{{U}}_S=\hat{\textbf{u}}+S(z^*-H/2)\textbf{{e}}_y$. The shear strength can be written as $S = U_b/H$, which is achieved by imposing streamwise velocity $\pm U_b/2$ at the top and bottom plates, respectively. We use the linear equation of state as $\rho=\rho_0\left[ 1-\beta_{\theta}\theta^* + \beta_s s^*) \right]$, where $\rho_0$ is the reference value for the density. The temperature $\theta^*$ and the salinity $s^*$ are also relative to the values of the reference state. $\beta_{\theta}$ is the thermal expansion coefficient and $\beta_s$ is the coefficient of the density increase due to salinity change. The governing equations for the incompressible velocity $\hat{\mathbf{u}}$ and the two scalar components then read, under the Oberbeck-Boussinesq assumption,
\begin{subequations}
\begin{eqnarray}
  \partial_t \hat{u}_i+\hat{u}_j\partial_j\hat{u}_i +U_{Sj}\partial_j\hat{u}_i 
     +\hat{u}_j\partial_jU_{Si} &=& -\partial_i p^* + \nu\partial_j^2\hat{u}_i 
       +g\delta_{iz^*}(\beta_{\theta}\theta^*-\beta_{s} s^*), \label{eqn:ns} \\
   \partial_t \theta^*+\hat{u}_j\partial_j\theta^*+U_{Sj}\partial_j\theta^* 
     &=& \kappa_{\theta}\partial_j^2\theta^*, \label{eqn:Tdiff} \\
   \partial_t \theta^*+\hat{u}_j\partial_js^*+U_{Sj}\partial_js^* 
     &=& \kappa_{s}\partial_j^2s^*, \label{eqn:Sdiff}\\
   \partial_j\hat{u}_j&=&0,\label{eqn:con}
\end{eqnarray}
\end{subequations}
in which $\hat{u}_i$ with $i=x^*,y^*,z^*$ are the three components of the perturbation velocity, $p^*$ is  kinematic pressure, $\nu$ is kinematic viscosity, $g$ is the gravitational acceleration, and $\kappa_\theta$ and $\kappa_s$ are the diffusivities of the temperature and the salinity, respectively.  The density has been absorbed in the pressure term of \eqref{eqn:ns}.

The temperature and the salinity are kept constant on the top and bottom plates, with the scalar differences across the fluid layer as $\Delta_\theta$ and $\Delta_s$, respectively. For the flow in the fingering regime, the top plate has higher temperature and salinity. Therefore, the convection flow is driven by the salinity field but stabilized by the temperature field. For the velocity $\hat{\mathbf{u}}$ on the two plates the stress-free condition is applied for the streamwise and spanwise components, and no penetration condition for the normal component, respectively. Periodic conditions are used in the two horizontal directions. All the variables are non-dimensionalized by the free-fall velocity $\sqrt{g\beta_s \Delta_s H}$, the domain height $H$ and the scalar differences $\Delta_\theta$ and $\Delta_s$. Then the non-dimensional equations are 
\begin{subequations}\label{eqn:wholend}
\begin{eqnarray}
	\partial_t u_i+u_j\partial_ju_i +\dfrac{(z-1/2)}{\sqrt{Ri}}\partial_y u_i +\dfrac{1}{\sqrt{Ri}}\delta_{iy}u_z  &=& -\partial_i p + \dfrac{\sqrt{Sc}}{\sqrt{Ra}}\partial_j^2u_i +\delta_{iz}(\Lambda\theta-s), \label{eqn:ns_nd} \\
	\partial_t \theta+u_j\partial_j\theta + \dfrac{(z-1/2)}{\sqrt{Ri}}\partial_y\theta
	&=& \dfrac{\sqrt{Sc}}{\sqrt{Ra}Pr}\partial_j^2\theta, \label{eqn:Tdiff_nd} \\
   	\partial_t s+u_j\partial_js + \dfrac{(z-1/2)}{\sqrt{Ri}}\partial_ys
   &=& \dfrac{1}{\sqrt{RaSc}}\partial_j^2s, \label{eqn:Sdiff_nd}\\
	\partial_ju_j&=&0,\label{eqn:con_nd}
\end{eqnarray} 
\end{subequations}
in which several control parameters are present. Throughout the current study we set the Prandtl number $Pr=\nu/\kappa_\theta=7$ and the Schmidt number $Sc=\nu/\kappa_s=700$, which are the typical values of seawater. The strength of the driving salinity gradient is measured by the salinity Rayleigh number $Ra=\beta_sg \Delta_s H^3/(\kappa_s\nu)$. The relative strength of the stabilizing temperature gradient is measured by the density ratio $\Lambda=\beta_\theta\Delta_\theta/\beta_s\Delta_s$. In addition, the background shear is characterized by the Richardson number $Ri = \beta_s g \Delta_s H / U_b^2$. Note that stronger shear corresponds to smaller $Ri$.

In the following sections we will first make a linear stability analysis for the governing equation \eqref{eqn:wholend} to identify the unstable parameter region and relevant unstable modes. Then we use our in-house DNS code to numerically solve the full nonlinear equations in the relevant parameter space. Some comparison will be made between the stability analyses and fully non-linear simulations.

\section{Linear stability analysis}\label{sec:stablity}

We now conduct a linear stability analysis for the non-dimensionalized governing equation \eqref{eqn:wholend}. The standard normal mode method is utilized. That is, the flow variables are decomposed as $\psi=\bar{\psi}+\psi'$, where $\bar{\psi}$ is the base state and $\psi'$ is the perturbation, respectively. For the current flow, the base state is chosen to be the background shear flow with a vertically linear distribution for both temperature and salinity, i.e. 
 \begin{equation}
	\bar{\textbf{u}}=\frac{z-1/2}{\sqrt{Ri}}\textbf{e}_y,\quad
	\bar{\theta}=z,\quad
	\bar{s}=z, \label{basestate}
\end{equation}
in which $z\in[0,1]$. In equation~\eqref{eqn:wholend} one can treat the velocity deviating from the uniform shear flow as the perturbation velocity. Then substituting the base state into the governing equation and neglect the high-order terms, the linearised equations read
\begin{subequations}\label{eqn:wholelin}
\begin{eqnarray}
	\partial_t u_i' + \frac{z-1/2}{\sqrt{Ri}} \partial_y u_i'
	   +\dfrac{1}{\sqrt{Ri}}\delta_{iy}u_z' 
	&=& - \partial_i p'+ \dfrac{\sqrt{Sc}}{\sqrt{Ra}}\partial_j^2u_i' 
	   +\delta_{iz}(\Lambda\theta'-s') , \label{eqn:ns_lin}  \\
	\partial_t \theta' +\frac{z-1/2}{\sqrt{Ri}} \partial_y \theta +u_z'
	&=& \frac{\sqrt{Sc}}{\sqrt{Ra}Pr}\nabla^2 \theta',  \label{eqn:Tdiff_lin} \\
	\partial_t s'+ \frac{z-1/2}{\sqrt{Ri}}\partial_y s'+u_z' 
	&=& \frac{1}{\sqrt{Ra Sc}}\nabla^2s' , \label{eqn:Sdiff_lin}\\
	\partial_j u_j'&=&0. \label{eqn:con_lin}
\end{eqnarray}
\end{subequations}
By taking the divergence of equation \eqref{eqn:ns_lin} and using the continuity condition \eqref{eqn:con_lin}, one obtains
\begin{equation}
	\frac{2}{\sqrt{Ri}} \partial_y u_z'=-\nabla^2 p' +\partial_z(\Lambda\theta' - s'). \label{eqn:temp_lin}
\end{equation}
To eliminate the pressure term, one can further take the $z$-derivative of equation \eqref{eqn:temp_lin} and minus it by the Laplacian of the $z$ component of equation \eqref{eqn:ns_lin}. A fourth-order differential equation for $u_z$ can then be obtained as
\begin{equation}
	\partial_t \nabla^2 u_z' = \frac{\sqrt{Sc}}{\sqrt{Ra}}\nabla^4 u_z'-\frac{z-1/2}{\sqrt{Ri}}  \partial_y \nabla^2 u_z'+\nabla^2_h(\Lambda \theta'- s'),
	\label{eqn:final_lin}
\end{equation}
in which $\nabla^2_h=\partial_x^2 + \partial_y^2$ is the Laplacian in the horizontal plane. Equations \eqref{eqn:Tdiff_lin}, \eqref{eqn:Sdiff_lin} and \eqref{eqn:final_lin} constitute an eigenvalue problem for three perturbation variables. We can numerically solve it by introduce a normal-mode solution as 
\begin{equation}
	\psi'(x,y,z,t)=\tilde{\psi}(z)\exp(ik_xx + ik_yy + \omega t), \label{equ:normalmode}
\end{equation}
in which $\psi$ stands for $u_z$, $\theta$ or $s$ and the tilde denotes the complex vertical shape functions. $k_x$ and $k_y$ are the real wave numbers in the horizontal plane. $\omega=\omega_r+i\omega_i$ is the complex temporal growth factor. To solve the eigen problem for different control parameters $(Ra, Ri, \Lambda)$ and wave numbers $(k_x, k_y)$, the Chebyshev polynomial expansion and the collocation method are adopted in the vertical direction.

\begin{figure}
	\centering
	\includegraphics[width=1\textwidth]{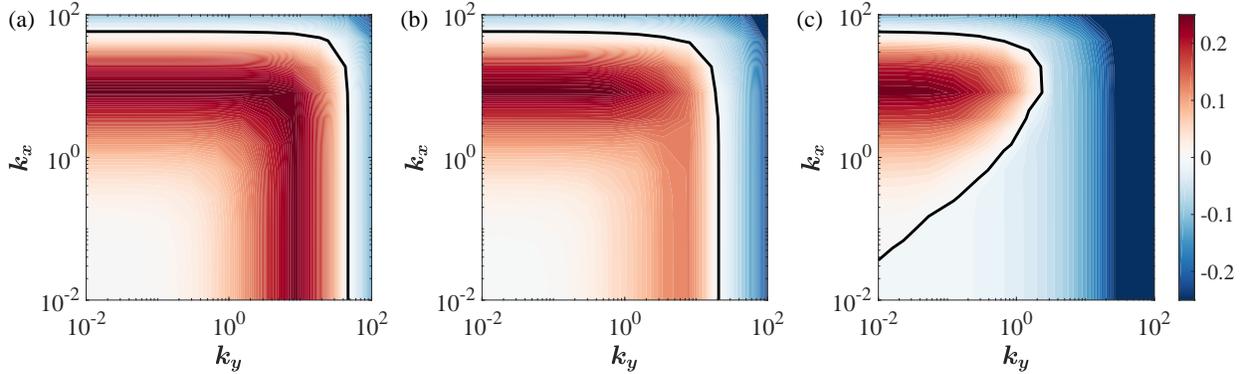}%
	\caption{The real part of the temporal growth factor $\omega_r$ in $(k_y,k_x)$ space for three cases with $Ra=10^7$, $\Lambda=1$: (a) $Ri=10^4$, (b) $Ri=10^2$ and (c) $Ri=1$. The black solid line denotes the neutral line ($\omega_r=0$). } 
	\label{fig:neutralline}
\end{figure}

In figure~\ref{fig:neutralline} we show the growth rate for modes with different wavenumbers. The Rayleigh number is fixed at $Ra=10^7$ and the density ratio at $\Lambda=1.0$. Three different Richardson numbers are considered, saying $Ri=10^4$, $10^2$, and $1$. The neutral curve with $\omega_r=0$ is also plotted for each set of parameters. In figure~\ref{fig:neutralline}a where the shear is very weak, there are very little differences of $\omega_r$ for different directions. The most unstable mode under this condition is supposed to be the salt finger instability. Detailed investigation shows that the large value of $\omega_r$ concentrates a bit more on the area where $k_x>k_y$, which is due to the streamwise shear. As will be shown later, although this anisotropy is very weak, it will lead to a big change for the arrangements of the salt fingers. This phenomenon becomes clearer when  $Ri$ decrease to $10^2$, as shown in figure \ref{fig:neutralline}b. The instability along the streamwise direction is faded a lot by the shear, but it is hardly affected in the spanwise direction. When $Ri$ equals unity in figure \ref{fig:neutralline}c, the most unstable mode totally reduces to the area where $k_y$ tends to zero, which leads to the two-dimensional salt sheets. This result is consistent with the previous linear stability analyses \citep{linden1974, smyth2007instability}. Note that even the smallest $Ri$ among our cases (i.e. $Ri=1$) is still larger than the criteria of KH instability ($Ri<1/4$), which is often considered in sheared finger studies \cite{smyth2011mixing, radko2011}. Furthermore, the vertical shear profile without inflection points also indicates that the two-dimensional salt sheets are the most unstable mode \cite{thangam1984}. Thus in the current study we expect no KH instability happens and focus on the interaction only between the primary salt-finger instability and the background shear.

\begin{figure}
	\centering
	\includegraphics[width=1\textwidth]{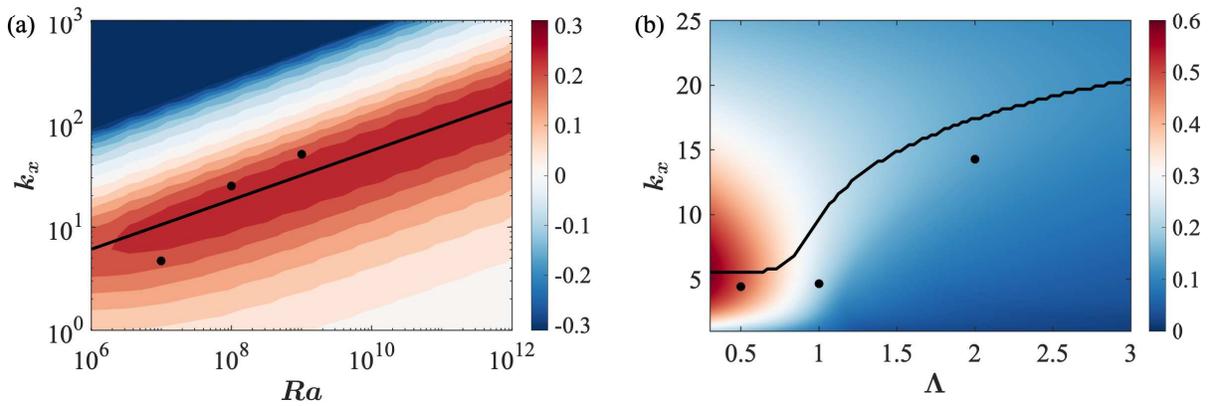}%
	\caption{The real part of the temporal growth factor $\omega_r$ as a function of (a) $Ra$ and $k_x$ and (b) $\Lambda$ and $k_x$. For both panels $k_y=0$. The black solid lines denote the fastest growing modes, and dots mark the spanwise length scales of the dominant structures in DNS results, respectively. The black line in (a) is given by linear fitting with a slope of $0.24$. The other control parameters are $Ri=10,~\Lambda=1$ in (a) and $Ri=10,~Ra=10^7$ in (b), respectively.} 
	\label{fig:fastmode}
\end{figure}
The streamwise shear has no effects on the pure spanwise mode $(k_y=0)$. One can easily deduce this conclusion by noticing that the terms related to $Ri$ disappear automatically when $k_y$ equals zero in equations~\eqref{eqn:Tdiff_lin}, \eqref{eqn:Sdiff_lin} and \eqref{eqn:final_lin}. As the result, when the salt sheets appear, they will no longer be affected by the shear. Specifically, the width of the salt sheets is related to $Ra$ and $\Lambda$, but independent of $Ri$. Thus we can choose any $Ri$ to find the performance of $\omega_r$ with other control parameters. In this way the salt sheets behave like the two-dimensional salt fingers. Figure \ref{fig:fastmode} shows $\omega_r$ as the functions of $(k_x, Ra)$ and $(k_x, \Lambda)$ with $k_y=0$ and $Ri=10$. The black solid lines denote the most unstable modes, and the black dots represent our DNS results, which will be discussed in the next section. We extract the width of the salt sheets $d$ from DNS and assume it to be the half wave length, i.e. $d=\pi / k_x$. Along the fastest growing line in figure \ref{fig:fastmode}a, the value of $\omega_r$ is almost unchanged when $Ra>10^7$, and $k_x$ varies linearly with $Ra$ in the logarithmic coordinates with a slope of $0.24$, indicating a scaling law of $d \sim k_x^{-1}\sim Ra^{-0.24}$. We calculated three cases with $Ra=10^7, 10^8, 10^9$ by DNS and the results are also close to this line. In figure \ref{fig:fastmode}b the fastest growing line has two different parts. According to the double diffusion theory~\citep{stern1960}, the salt finger instability happens when $1<\Lambda<\kappa_\theta/\kappa_{s}=100$. In this part of the line, $k_x$ increase monotonously with $\Lambda$ with a decreasing growth rate. For $\Lambda<1$, the flow is susceptible to the gravitationally unstable convection, and the most unstable wave number remains nearly unchanged. The DNS results cover $\Lambda=0.5, 1 ,2$. As will be shown later, the salt fingers still appear in the $\Lambda=0.5$ case since the boundary layers make the bulk density ratio larger than the preset value. Compared with the stability analysis, the DNS results have some deviation, but the overall performance is consistent. In the following passage we will show the DNS results in detail. 

\section{Three-dimensional direct numerical simulation}\label{sec:dns}

\subsection{Numerical settings}

We now turn to the fully nonlinear simulations. The numerical method will be briefly described first, then the numerical results will be discussed. We use our well developed code to solve the governing equation~\ref{eqn:wholend}. The code employs a fraction of time step method with the finite-difference scheme. An advantage of this code is the use of the multi-grid method, namely, a refined mesh is established for the salinity field because of its relatively small diffusivity, while the other variables are solved on the basic mesh. The code has been widely used in our previous DDC cases and other turbulent simulations~\cite{ostilla2014,ddcjfm2016}. Initially both the temperature and salinity increase linearly from bottom to top, while velocity $\mathbf{u}$ is set to zero. Besides, small perturbations are added to trigger the flow. All the simulations have been run until the statistically steady state is reached, and the flow fields and statistical results are extracted from the statistically steady stage. 

Details of the parameter setting are shown in table~\ref{tab:case1} in the Appendix. Specifically, all cases are divided into five groups by the different Rayleigh number $Ra$ and density ratio $\Lambda$. The first three groups have the same Rayleigh number $Ra=10^7$ and different density ratios $\Lambda=0.5$, $1$, and $2$, respectively. Then we keep $\Lambda=1$ and increase $Ra$ to $10^8$ and $10^9$, as shown in the last two groups. Within each group, the Richardson number $Ri$ gradually varies in a large range over five or six orders of magnitude. 

\subsection{Flow morphology and transport properties at $Ra=10^7$}\label{sec:fixra}

\begin{figure}
	\centering
	\includegraphics[width=1\textwidth]{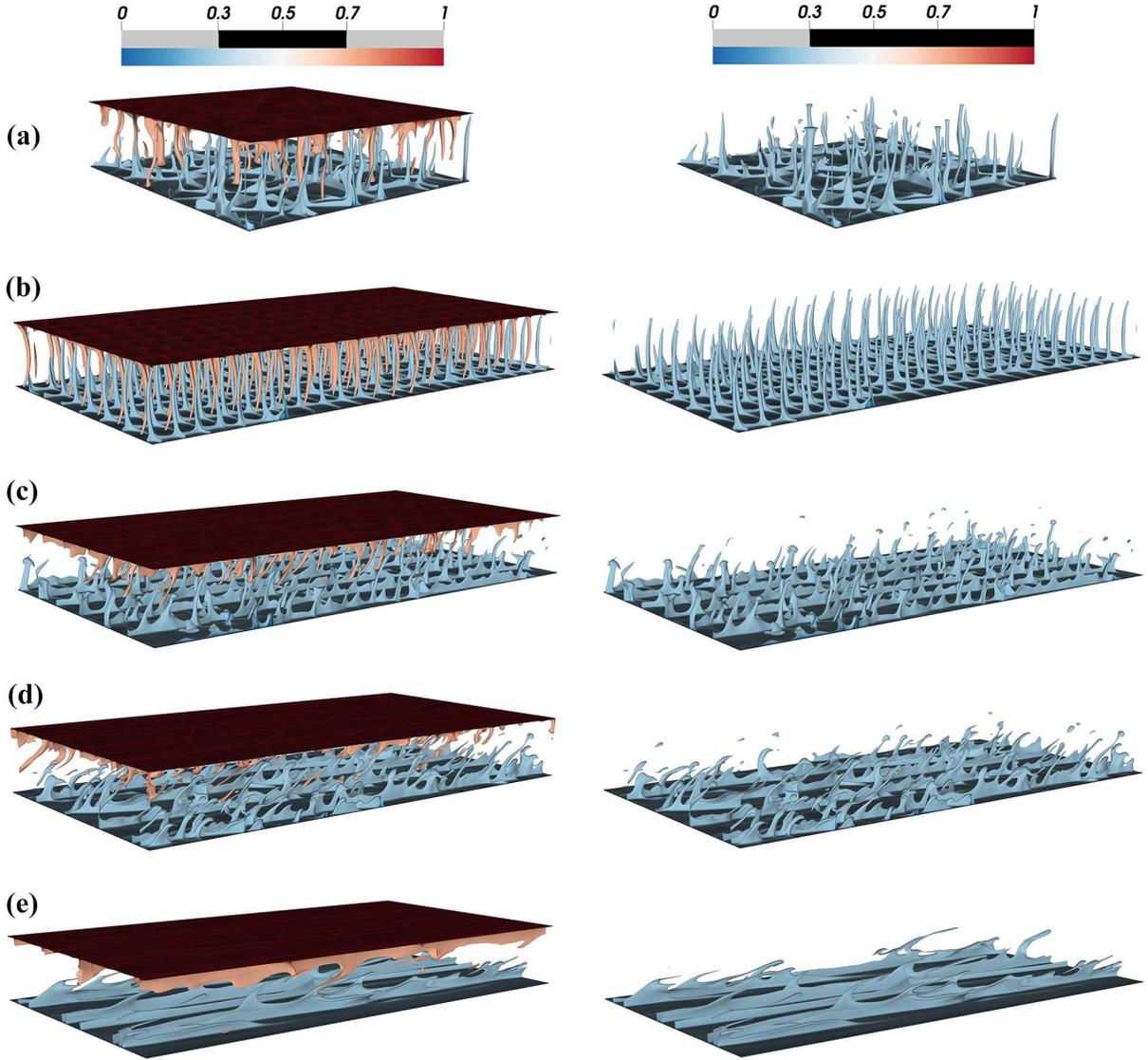}%
	\caption{Three dimensional volume rendering of the instantaneous salinity field with (a) no background shear, (b) $Ri=10^4$, (c) $Ri=10^3$, (d) $Ri=10^2$ and (e) $Ri=10$. The color is determined by the salinity in two end values (left) and single end value (right). The other control parameters all read $Ra=10^7$, $\Lambda=1$. } 
	\label{fig:3dfinger}
\end{figure}

In this section we focus on the cases with $Ra=10^7$ and varying $\Lambda$ and $Ri$. We first look at the morphology change of finger structures as the shear enhances for $\Lambda=1$. Our results suggest that these qualitative behaviors are similar for different $Ra$ and $\Lambda$. Figure~\ref{fig:3dfinger} presents three-dimensional volume rendering of instantaneous salinity fields for five different shear rates. The left column shows the structures with salinity smaller than $0.3\Delta_s$ and larger than $0.7\Delta_s$, which include the finger structures both ascending from the bottom plate and descending from the top plate. The right column only shows the finger structures growing from the bottom plate with salinity smaller than $0.3\Delta_s$. In the absence of shear, as shown in figure~\ref{fig:3dfinger}a, the salt fingers originated from both plates are arranged disorderly in their horizontal locations, which is similar to the previous simulations~\cite{ddcjfm2015,ddcjfm2016}. Due to the relatively low $Ra$, most of the salt fingers reach the opposite plate. Interestingly, a weak shear already has profound effects on the horizontal arrangement of the fingers. For $Ri=10^4$, saying the shear velocity is only one percent of the free-fall velocity, the salt fingers are still vertically oriented but their horizontal locations become very well organized. Fingers rooted from the same plate form nearly straight lines which are along the streamwise direction. Meanwhile, ascending fingers and descending ones appear alternatively in the spanwise direction. For $Ri=10^3$ the salt fingers exhibit distinct tilting towards the shearing direction and sheet-like structures along the streamwise direction emerge at the roots of fingers near the plates. As $Ri$ decreases further the sheet-like structures are more clear. For $Ri=10$, namely the strongest shear considered here, the dominant structures become nearly two-dimensional salt sheets, as shown in figure~\ref{fig:3dfinger}e. It should be noted that $Ri=10$ is close to the typical value measured in the Ocean~\cite{gregg1987}. If $Ri$ continues to decrease, secondary instability may develop for the salt-sheet structures~\cite{smyth2011c}.

\begin{figure}
	\centering
	\includegraphics[width=1\textwidth]{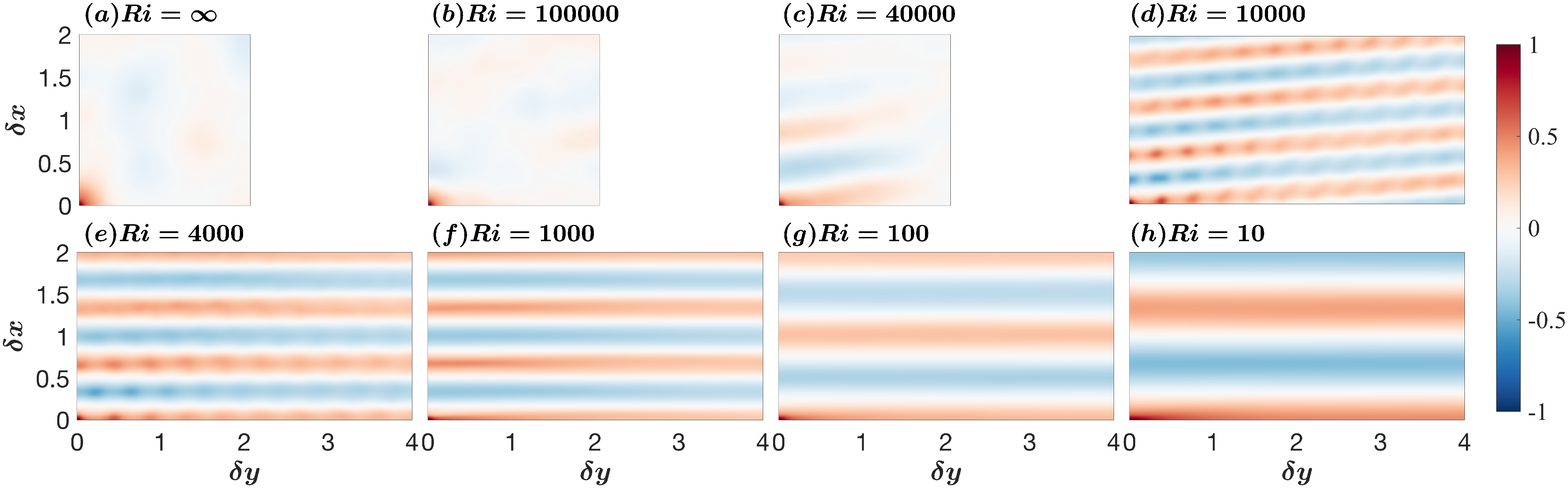}%
	\caption{The auto-correlation coefficient $C_s$ of salinity  for the streamwise separation $\delta_y$ and the spanwise separation  $\delta_x$ in the vertical mid-plane. Different shear strengths are set as unsheared for (a) and $Ri=10^5, 4\times10^4, 10^4, 4\times10^3, 10^3, 10^2, 10$ for (b) - (h), respectively. The other control parameters all read $Ra=10^7$, $\Lambda=1$.} 
	\label{fig:sa_autocorr}
\end{figure}

Above discussions indicate that as the shear strengthens first the horizontal pattern of fingers is altered, then fingers are tilted, and finally fingers are replaced by salt sheets along the streamwise direction. To quantitatively investigate such behaviors, we calculate the auto-correlation coefficient $C_s$ of salinity in the horizontal directions as
\begin{equation}\label{eqn:sautocorr}
	C_s(\delta_x,\delta_y) = \frac{\langle(s(x,y)-\mu_s)(s(x+\delta_x,y+\delta_y)-\mu_s)\rangle_h}{\sigma^2_{s}(x,y)},
\end{equation}
in which $\mu_s$ and $\sigma_s$ are the mean and the standard deviation of salinity over the given horizontal plane and time, respectively. Hereafter, $\langle~\rangle_h$ stands for the temporal and spatial average over a horizontal plane. By definition $-1 \le C_s \le 1$ and $C_s(0,0)=0$. The coefficient $C_s$ is computed for the cases with $Ra=10^7$ and $\Lambda=1$ over the horizontal plane at the height $0.2H$. The results are shown in figure~\ref{fig:sa_autocorr}. For the case without shear, the auto-correlation coefficient is isotropic in the $(\delta x, \delta y)$ plane, as shown in panel a. This is expectable since fingers randomly distribute in the horizontal directions without any preference. When a weak shear is applied, the correlation immediately becomes stronger along the line with a small angle to the $\delta y$ direction, saying close to streamwise direction. For $Ri=10^4$ an organized pattern emerges in the contours of $C_s$, see figure~\ref{fig:sa_autocorr}d. Patches with large positive and negative values appear with a regular spacing. Compared with figure~\ref{fig:3dfinger}b, it is obvious that the organized pattern of $C_s$ corresponds to the regularly distributed fingers. As the shear further strengthens, the organized pattern disappears and the auto-correlation coefficient is dominated by a strong strip along the $\delta y$ direction, indicating streamwise oriented sheet-like structures.

\begin{figure}
	\centering
	\includegraphics[width=0.6\textwidth]{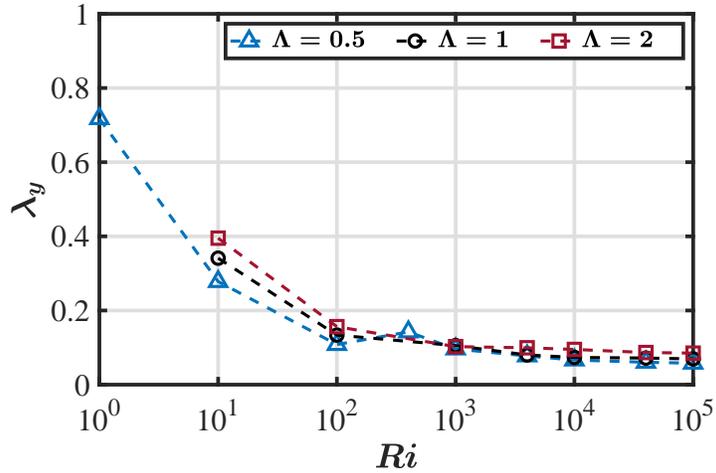}%
	\caption{The streamwise correlation length scale determined from the auto-correlation function versus the shear strength. For all cases $Ra$ is fixed at $10^7$.} 
	\label{fig:scalefixra}
\end{figure}

Some typical length scales can be extracted from the auto-correlation coefficient $C_s$. From the function $C_s(\delta x=0, \delta y)$ one can use a parabola to fit the curve close to the original point, then a length scale $\lambda_y$ can be defined as the intersection location of this parabola and the $\delta y$-axis. Clearly $\lambda_y$ is related to the decreasing rate of $C_s$ along the $\delta y$-axis and indicates the correlation length of the salinity field along the streamwise direction. The dependence of $\lambda_y$ on the shear strength is plotted in figure~\ref{fig:scalefixra}. As the shear becomes stronger, saying $Ri \rightarrow 0$, the streamwise correlation length first keeps constant and then increases for all three density ratios considered. This corresponds to the fact that fingers becomes organized at weak shear and then are replaced by sheets when shear is strong enough. Another length scale is the spanwise spacing $d$ of the salt-finger or salt-sheet structures, which can be determined by the value of $\delta x$ at the first negative minimum of the curve $C_s(\delta x, \delta y=0)$. The results are summarized in table \ref{tab:case1}. When the salt sheets totally replace the finger structure, $d$ is assumed to be the width of the sheets and independent of $Ri$, which is compared with the linear results in figure \ref{fig:fastmode}b. It can be seen that the linear stability analysis captures the variation trend of this spanwise spacing scale.

\begin{figure}
\centering
\includegraphics[width=\textwidth]{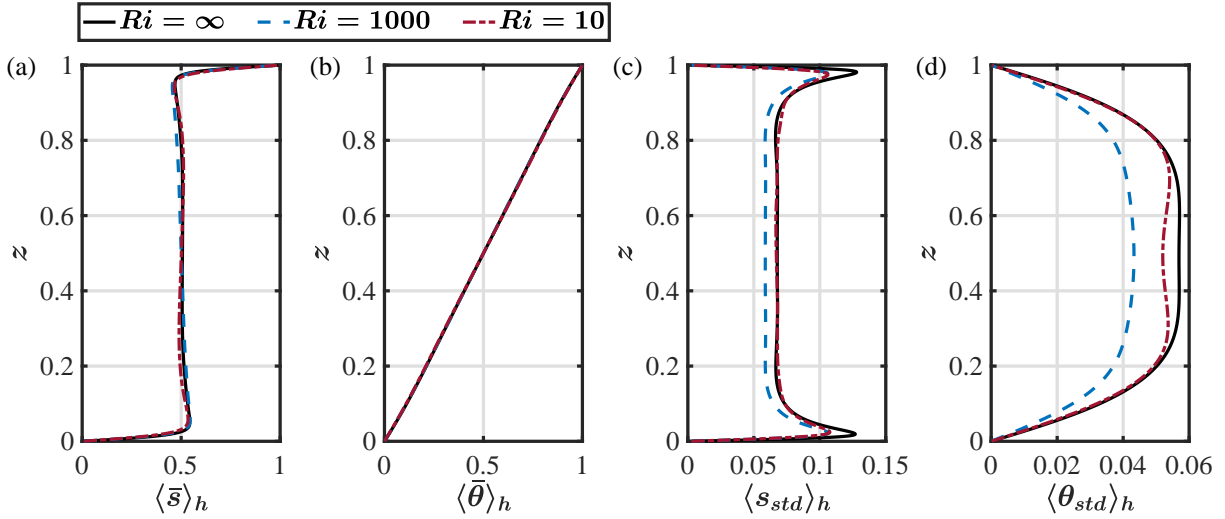}%
\caption{The mean profiles of (a) salinity, (b) temperature, (c) standard deviation of salinity and (d) standard deviation of temperature. The bar and the bracket stand for the temporal and horizontal average value, respectively. In each panel three cases with $Ra=10^7, \Lambda=1$ and different shear strengths are shown.} 
\label{fig:prof}
\end{figure}
\begin{figure}
	\centering
	\includegraphics[width=\textwidth]{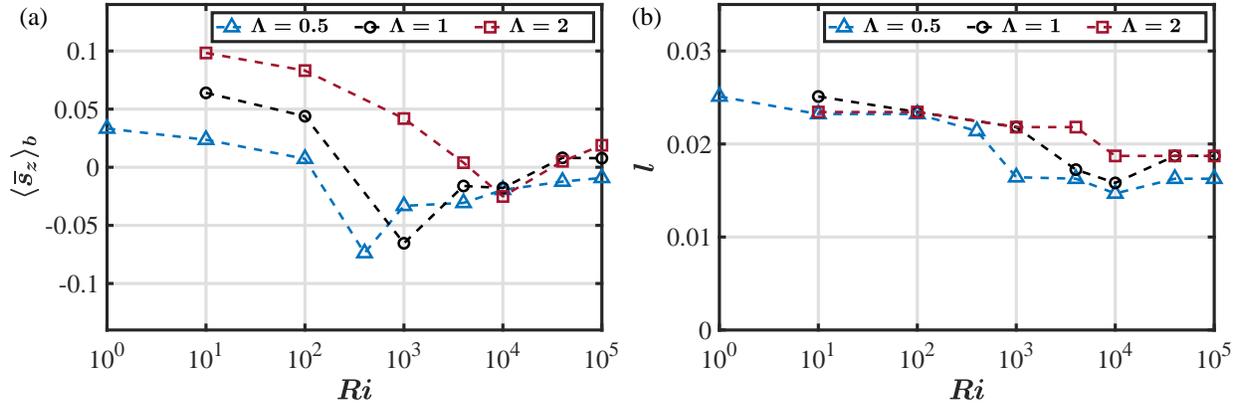}%
	\caption{(a) The central salinity gradient $\langle \partial_z\bar{s} \rangle_b$ vs $Ri$. (b) The salinity boundary layer thickness $l$ vs $Ri$. For all cases $Ra$ is fixed at $10^7$.} 
	\label{fig:prof2}
\end{figure}
We now turn to the mean profiles of scalar fields. Figure~\ref{fig:prof} shows the vertical profiles of salinity and temperature averaged both in time and over the horizontal plane. Three cases are shown for $(Ra,\Lambda)=(10^7,1)$ and $Ri=\infty$, $1000$, and $10$, respectively. Due to the large molecular diffusivity, the temperature profiles are very close to the linear one, as shown in figure \ref{fig:prof}b. The salinity field has distinct boundary layer and bulk regions, see figure~\ref{fig:prof}a. As the shear becomes stronger, the vertical gradient of mean salinity around the mid height changes from slightly positive to negative, and then becomes positive again. We calculate the salinity vertical gradient averaged in the middle bulk area for each case, i.e. $\langle \partial_z\bar{s} \rangle_b=(\langle \bar{s} \rangle_{z=0.6}-\langle \bar{s} \rangle_{z=0.4})/0.2$. The results are summarized in table~\ref{tab:case1} and plotted in figure \ref{fig:prof2}a. The negative mean salinity gradient at domain center only happens for weak shear cases in which salt fingers becomes very organized. The profiles for the standard deviation of salinity and temperature are shown in figures~\ref{fig:prof}c and \ref{fig:prof}d. For the weak shear case with negative mean salinity gradient at the center, the fluctuations of both scalars are also weaker compared to other cases. The salinity boundary layer thickness $l$ can be calculated by the maximum value points in figure \ref{fig:prof}c, and are summarized in table~\ref{tab:case1} and plotted in figure \ref{fig:prof2}b. For the weak shear cases $l$ nearly keeps constant, while it becomes thicker visibly under the strong shear.

\begin{figure}
	\centering
	\includegraphics[width=1\textwidth]{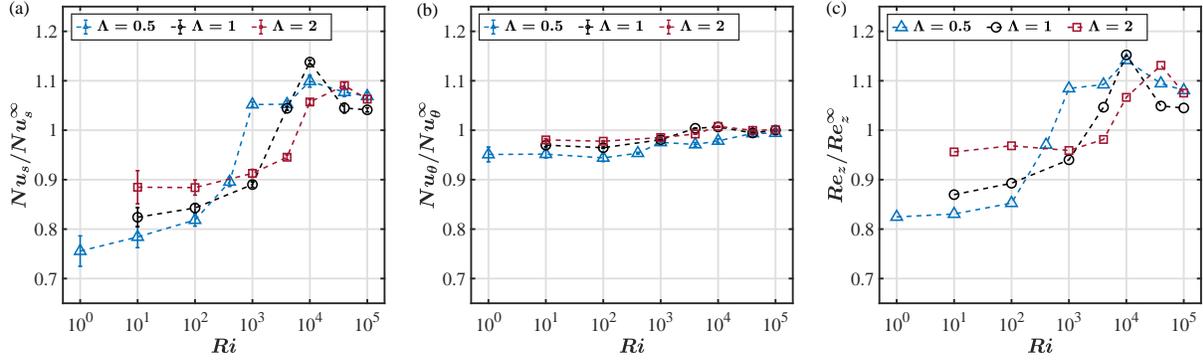}%
	\caption{The global fluxes for different shear strengths normalized by the values of no-shear cases. (a) Salinity Nusselt number, (b) heat Nusselt number, and (c) Reynolds number. For all cases $Ra$ is fixed at $10^7$.} 
	\label{fig:fluxfixra}
\end{figure}
The change of structure morphology affects the vertical transport property of the system. Three non-dimensional numbers are used to measure the heat, salt, and momentum fluxes, saying the two Nusselt number and the Reynolds number defined as
\begin{equation}
	Nu_\theta=\frac{\langle u_z \theta \rangle_h - \kappa_\theta\partial_z\langle\theta\rangle_h} 
	{\kappa_\theta \Delta_\theta H^{-1}}, \quad \quad
	Nu_s=\frac{\langle u_z s \rangle_h - \kappa_s\partial_z\langle s \rangle_h} 
	{\kappa_s \Delta_s H^{-1}}, \quad \quad
	Re_z=\frac{u_z^{rms} H }{\nu}.
\end{equation}
Note that we calculate the Reynolds number by the root-mean-square value of the vertical velocity $u_z$ to measuring the vertical momentum transport. These quantities are plotted in figure~\ref{fig:fluxfixra}. Interestingly, both $Nu_s$ and $Re_z$ first increase and then decrease as $Ri$ gradually decreases. That is, for weak shear, both salinity transfer and vertical motion are enhanced. While they are suppressed for stronger shear. The enhancement of these two quantities are caused by the organized fingers which generate a more efficient transport than the no-shear case. Similar phenomena have also be reported in other systems~\cite{chong2017}. The heat Nusselt number, however, only exhibits a very weak variation for different shear strengths. This is expectable since in the current setting heat diffuses much quicker than salt. For the relatively low $Ra=10^7$, diffusion dominates the heat transport and shear only has a minor influence.

\subsection{Influences of different Rayleigh numbers}\label{sec:fixdr}
\begin{figure}
	\centering
	\includegraphics[width=1\textwidth]{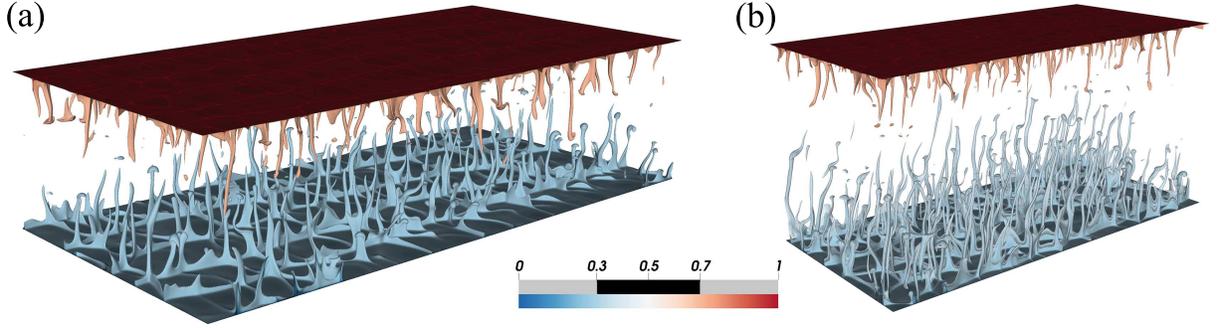}%
	\caption{Three dimensional volume rendering of the instantaneous salinity field with (a) $Ra=10^8$ and (b) $Ra=10^9$. The other control parameters read $Ri=10^4$, $\Lambda=1$. } 
	\label{fig:highra3dfinger}
\end{figure}
\begin{figure}
	\centering
	\includegraphics[width=0.6\textwidth]{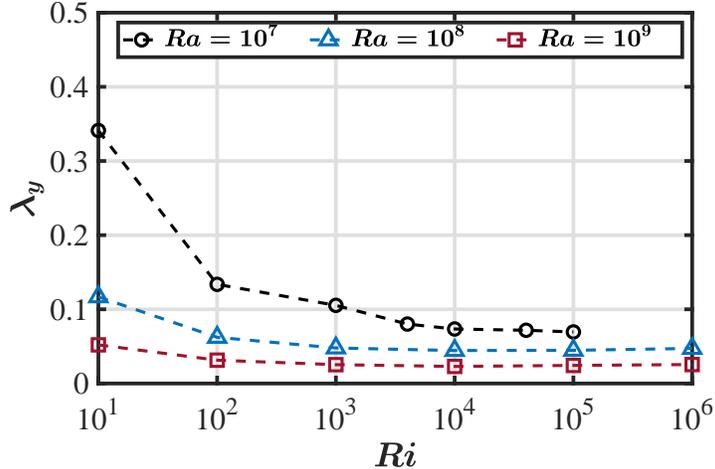}%
	\caption{The streamwise correlation length scale determined from the auto-correlation function versus the shear strength. $\Lambda$ is fixed at $1$.} 
	\label{fig:scalefixdr}
\end{figure}

We now investigate the influences of the Rayleigh number. $\Lambda$ is fixed at unity and three Rayleigh numbers are considered, saying $Ra=10^7$, $10^8$, and $10^9$. As discussed in the previous section, for all the $Ra$ and $\Lambda$ considered here, the flow morphology shows similar transition as those shown in figure~\ref{fig:3dfinger}. Here, figure \ref{fig:highra3dfinger} presents the three dimensional volume rendering of the instantaneous salinity field with $Ra=10^8$ and $10^9$ for $Ri=10^4$. It should be pointed out that the salt fingers have similar strength between those ascending from the bottom and those descending from the top. The regularity of the flow structures weakens with the increase of $Ra$ for the same shear strength. Moreover, the salt fingers become more turbulent and quickly lose their salinity anomaly before reaching the opposite boundary. In figure~\ref{fig:scalefixdr} we plot the behaviours of $\lambda_y$ versus $Ri$ for the three different $Ra$. For all Rayleigh numbers the streamwise correlation length increases as shear becomes stronger, indicating the formation of sheet-like structures. However, $\lambda_y$ is smaller for larger $Ra$. Moreover, even for the strongest shear with $Ri=10$ the increment of $\lambda_y$ with respect to the no-shear case is rather small, i.e. the coherence of the salt-sheet structures is still quite weak. For higher Rayleigh number as in the Ocean, the salt-sheet structures may only develop at even stronger shear. 

\begin{figure}
	\centering
	\includegraphics[width=1\textwidth]{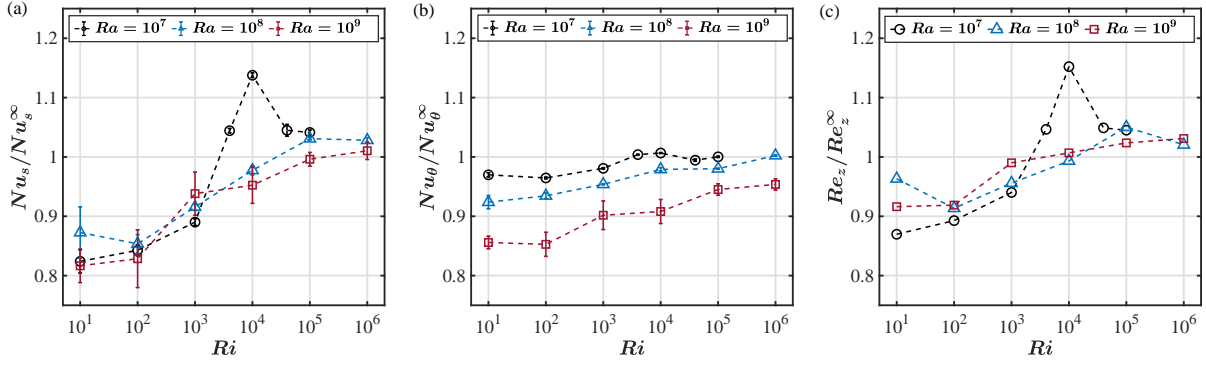}%
	\caption{The global fluxes for different shear strengths normalized by the values of no-shear cases. (a) Salinity Nusselt number, (b) heat Nusselt number, and (c) Reynolds number. $\Lambda$ is fixed at $1$.} 
	\label{fig:fluxfixdr}
\end{figure}

Different behaviors of the correlation length scales also reflect themselves in the variations of fluxes. In figure~\ref{fig:fluxfixdr} we plot the two Nusselt numbers and the Reynolds number, all normalized by the values of the no-shear cases. For $Ra=10^7$, both $Nu_s$ and $Re_z$ are enhanced considerably for weak shear. However, this enhancement is smaller as $Ra$ becomes larger. For $Ra=10^9$ there is effectively no enhancement in $Nu_s$. This can be attributed to the less coherence of salt-finger structures along the vertical direction at high $Ra$. At the strongest shear studied here, $Nu_s$ can be almost $20\%$ less than that of no-shear case. For large $Ra$, shear can also reduce the heat flux significantly. With strong buoyancy driving force, convection starts to play an apparent role in the dynamics of temperature component, instead of the nearly conduction state for $Ra=10^7$. And then shear can affect the heat flux.  

\section{Conclusions} \label{sec:con}

In summary, the salt-finger DDC with a uniform shear is studied by both linear stability analysis and fully nonlinear simulation. We demonstrate that a weak shear can already alter the flow morphology significantly. Specifically, salt fingers distribute regularly when a weak shear is applied. As the background shear further increases, the dominant structures change to salt sheets parallel to the shearing velocity, and the structures become increasingly correlated along the streamwise direction. For the current flow bounded by two plates from top and bottom, linear stability analyses reveal similar trend: the most unstable modes become two-dimensional as shear is enhanced. Interestingly, the scales of the most unstable modes given by linear theory are quite close to those measured from the nonlinear simulations of fully developed flows.

The effects of shear on global transport is rather complex. For all combinations of Rayleigh number and density ratio, the temperature Nusselt number $Nu_T$ decreases as shear becomes stronger. At relatively low Rayleigh number $Ra=10^7$, the decrement of $Nu_T$ is minor due to the fact that the thermal field is effectively in the conductive state. For larger $Ra$, however, $Nu_T$ can decrease by over $10\%$ compared to the case without shear. The salinity Nusselt number, on the other hand, first increases and then decreases as the Richardson number $Ri$ decreases. The enhancement of $Nu_S$ is related to the regularization of salt-finger distribution under weak shear. The magnitude of this enhancement is smaller for larger $Ra$. Moreover, $Nu_S$ decreases by about $20\%$ for the strongest shear considered here. 

Our results not only reveal interesting results about the fingering DDC with weak shear, but also have important implications for oceanic DDC. Especially, even a weak shear can notably alter the heat and salt fluxes in the vertical direction. Moreover, it affects the two fluxes in different ways, therefore also the density flux ratio is changed.  

\bigskip

\noindent{\it Acknowledgements:} This work is supported by the Major Research Plan of National Natural Science Foundation of China for Turbulent Structures under the grants 91852107 and 91752202.

\appendix

\section{Numerical details}

In this appendix we provide the numerical details and key responses for all the simulated cases.
\begin{table}
	\setlength{\tabcolsep}{2pt}
	\begin{ruledtabular}
	\begin{tabular}{lcccccccccccccc}
		$Ra$&$\Lambda$&$Ri$&$L_x$&$L_y$&$N_x(m_x)$&$N_y(m_y)$ &$N_z(m_z)$&$\lambda_y$&$d$&$\langle \partial_z\bar{s} \rangle_b$&$l$&$Nu_s$&$Nu_\theta$&$Re_z$\\ \hline \\
		$1\times10^7$&$0.5$&$1\times10^0$&$4$&$8$&$192(4)$&$384(5)$&$128(2)$&$0.718$&$0.711$&$0.033$&$0.025$&$25.6$&$1.26$&$0.77$\\
		$1\times10^7$&$0.5$&$1\times10^1$&$4$&$8$&$192(4)$&$384(5)$&$128(2)$&$0.278$&$0.680$&$0.024$&$0.023$&$26.5$&$1.26$&$0.77$\\
		$1\times10^7$&$0.5$&$1\times10^2$&$4$&$8$&$192(4)$&$384(5)$&$128(2)$&$0.108$&$0.680$&$0.007$&$0.023$&$27.7$&$1.25$&$0.79$\\
		$1\times10^7$&$0.5$&$4\times10^2$&$4$&$8$&$192(4)$&$384(5)$&$128(2)$&$0.142$&$0.346$&$-0.074$&$0.021$&$30.3$&$1.26$&$0.90$\\
		$1\times10^7$&$0.5$&$1\times10^3$&$4$&$4$&$256(4)$&$256(4)$&$128(3)$&$0.096$&$0.354$&$-0.033$&$0.016$&$35.6$&$1.29$&$1.01$\\
		$1\times10^7$&$0.5$&$4\times10^3$&$4$&$4$&$256(4)$&$256(4)$&$128(2)$&$0.078$&$0.318$&$-0.031$&$0.016$&$35.6$&$1.26$&$1.02$\\
		$1\times10^7$&$0.5$&$1\times10^4$&$4$&$4$&$256(4)$&$256(4)$&$128(2)$&$0.067$&$0.299$&$-0.020$&$0.015$&$37.2$&$1.30$&$1.06$\\
		$1\times10^7$&$0.5$&$4\times10^4$&$4$&$4$&$256(4)$&$256(4)$&$128(2)$&$0.060$&$0.381$&$-0.012$&$0.016$&$36.4$&$1.32$&$1.02$\\
		$1\times10^7$&$0.5$&$1\times10^5$&$4$&$4$&$192(4)$&$192(4)$&$128(2)$&$0.058$&$0.445$&$0.009$&$0.016$&$36.2$&$1.32$&$1.01$\\ 
		$1\times10^7$&$0.5$&$\infty$&$4$&$4$&$192(4)$&$192(4)$&$128(2)$&$0.059$&$0.904$&$-0.011$&$0.018$&$33.8$&$1.32$&$0.93$\\[0.2cm]
			
		$1\times10^7$&$1$&$1\times10^1$&$4$&$8$&$288(3)$&$288(4)$&$144(2)$&$0.341$&$0.674$&$0.064$&$0.025$&$26.7$&$1.14$&$0.67$\\ 
		$1\times10^7$&$1$&$1\times10^2$&$4$&$8$&$288(3)$&$288(4)$&$144(2)$&$0.134$&$0.512$&$0.044$&$0.023$&$27.3$&$1.13$&$0.69$ \\ 
		$1\times10^7$&$1$&$1\times10^3$&$4$&$8$&$288(3)$&$288(4)$&$144(2)$&$0.105$&$0.336$&$-0.065$&$0.022$&$28.8$&$1.15$&$0.73$ \\ 	
		$1\times10^7$&$1$&$4\times10^3$&$4$&$8$&$288(3)$&$288(4)$&$144(2)$&$0.080$&$0.331$&$-0.016$&$0.017$&$33.8$&$1.18$&$0.81$ \\ 	
		$1\times10^7$&$1$&$1\times10^4$&$4$&$8$&$288(3)$&$288(4)$&$144(2)$&$0.074$&$0.285$&$-0.018$&$0.016$&$36.8$&$1.18$&$0.89$ \\ 
		$1\times10^7$&$1$&$4\times10^4$&$4$&$4$&$288(3)$&$288(3)$&$144(2)$&$0.072$&$0.414$&$0.008$&$0.019$&$33.8$&$1.17$&$0.81$ \\ 	
		$1\times10^7$&$1$&$1\times10^5$&$4$&$4$&$288(3)$&$288(3)$&$144(2)$&$0.070$&$0.433$&$0.008$&$0.019$&$33.7$&$1.17$&$0.81$ \\ 	
		$1\times10^7$&$1$&$\infty$&$4$&$4$&$288(3)$&$288(3)$&$144(2)$&$0.075$&$0.674$&$0.012$&$0.019$&$32.3$&$1.17$&$0.78$ \\ 	
	\end{tabular}
	\caption{Summary of the simulations. For all simulations, we keep the length in the vertical direction $L_z=1$, the Prandtl number $Pr=7$ and the Schmidt number $Sc=700$ . Columns from left to right are: the Rayleigh number of salinity, the density ratio, the Richardson number, lengths in the spanwise, steamwise, and normal directions, resolutions in the spanwise, steamwise, normal directions (with refinement factors for multiple resolutions), the streamwise correlation length scale, the spanwise spacing scale, the bulk-averaged salinity vertical gradient, the salinity boundary layer thickness, the salinity and temperature Nusselt numbers and the vertical Reynolds number.}
	\label{tab:case1}
	\end{ruledtabular}
\end{table}

\begin{table}
	\setlength{\tabcolsep}{2pt}
	\begin{ruledtabular}
	\begin{tabular}{lcccccccccccccc}
		$Ra$&$\Lambda$&$Ri$&$L_x$&$L_y$&$N_x(m_x)$&$N_y(m_y)$ &$N_z(m_z)$&$\lambda_y$&$d$&$\langle \partial_z\bar{s} \rangle_b$&$l$&$Nu_s$&$Nu_\theta$&$Re_z$\\ \hline \\
		$1\times10^7$&$2$&$1\times10^1$&$4$&$8$&$288(3)$&$288(4)$&$144(2)$&$0.395$&$0.220$&$0.098$&$0.023$&$27.4$&$1.07$&$0.62$  \\ 
		$1\times10^7$&$2$&$1\times10^2$&$4$&$8$&$288(3)$&$288(4)$&$144(2)$&$0.156$&$0.215$&$0.083$&$0.023$&$27.4$&$1.07$&$0.63$ \\ 
		$1\times10^7$&$2$&$1\times10^3$&$4$&$8$&$288(3)$&$288(4)$&$144(2)$&$0.103$&$0.326$&$0.042$&$0.022$&$28.3$&$1.07$&$0.62$ \\ 
		$1\times10^7$&$2$&$4\times10^3$&$4$&$8$&$288(3)$&$288(4)$&$144(2)$&$0.100$&$0.280$&$0.004$&$0.022$&$29.3$&$1.08$&$0.64$ \\ 	
		$1\times10^7$&$2$&$1\times10^4$&$4$&$8$&$288(3)$&$288(4)$&$144(2)$&$0.095$&$0.252$&$-0.025$&$0.019$&$32.8$&$1.10$&$0.69$ \\ 	
		$1\times10^7$&$2$&$4\times10^4$&$4$&$4$&$288(3)$&$288(3)$&$144(2)$&$0.087$&$0.220$&$0.005$&$0.019$&$33.8$&$1.09$&$0.74$ \\ 	
		$1\times10^7$&$2$&$1\times10^5$&$4$&$4$&$288(3)$&$288(3)$&$144(2)$&$0.085$&$0.303$&$0.019$&$0.019$&$32.9$&$1.09$&$0.70$ \\ 
		$1\times10^7$&$2$&$\infty$&$4$&$4$&$288(3)$&$288(3)$&$128(2)$&$0.097$&$1.243$&$0.040$&$0.021$&$31.0$&$1.09$&$0.65$ \\[0.2cm]
			
		$1\times10^8$&$1$&$1\times10^1$&$4$&$8$&$384(3)$&$768(3)$&$192(2)$&$0.117$&$0.127$&$0.075$&$0.011$&$55.6$&$1.25$&$1.95$  \\ 
		$1\times10^8$&$1$&$1\times10^2$&$4$&$8$&$384(3)$&$768(3)$&$192(2)$&$0.062$&$0.373$&$0.051$&$0.011$&$54.4$&$1.26$&$1.85$  \\ 
		$1\times10^8$&$1$&$1\times10^3$&$4$&$8$&$384(3)$&$768(3)$&$192(2)$&$0.048$&$0.352$&$0.012$&$0.011$&$58.4$&$1.29$&$1.93$  \\ 
		$1\times10^8$&$1$&$1\times10^4$&$4$&$8$&$384(3)$&$768(3)$&$192(2)$&$0.044$&$0.217$&$-0.001$&$0.010$&$62.3$&$1.33$&$2.01$  \\ 
		$1\times10^8$&$1$&$1\times10^5$&$4$&$4$&$480(3)$&$480(3)$&$192(2)$&$0.045$&$0.357$&$0.002$&$0.009$&$65.7$&$1.33$&$2.12$  \\ 
		$1\times10^8$&$1$&$1\times10^6$&$4$&$4$&$384(3)$&$384(3)$&$192(2)$&$0.047$&$0.401$&$0.001$&$0.009$&$65.5$&$1.36$&$2.06$  \\ 
		$1\times10^8$&$1$&$\infty$&$4$&$4$&$384(3)$&$384(3)$&$192(2)$&$0.052$&$0.589$&$0.001$&$0.009$&$63.8$&$1.35$&$2.02$ \\[0.2cm]
			
		$1\times10^9$&$1$&$1\times10^1$&$1$&$2$&$240(4)$&$480(4)$&$288(3)$&$0.052$&$0.062$&$0.121$&$0.005$&$110$&$1.51$&$5.04$ \\ 
		$1\times10^9$&$1$&$1\times10^2$&$1$&$2$&$240(4)$&$480(4)$&$288(3)$&$0.031$&$0.061$&$0.059$&$0.005$&$112$&$1.51$&$5.05$ \\ 
		$1\times10^9$&$1$&$1\times10^3$&$1$&$2$&$240(4)$&$480(4)$&$288(3)$&$0.025$&$0.069$&$0.038$&$0.005$&$126$&$1.59$&$5.45$ \\ 
		$1\times10^9$&$1$&$1\times10^4$&$1$&$2$&$240(4)$&$480(4)$&$288(3)$&$0.023$&$0.068$&$0.019$&$0.005$&$128$&$1.60$&$5.54$ \\ 
		$1\times10^9$&$1$&$1\times10^5$&$1$&$2$&$240(4)$&$480(4)$&$288(3)$&$0.024$&$0.077$&$0.009$&$0.004$&$134$&$1.67$&$5.63$ \\ 
		$1\times10^9$&$1$&$1\times10^6$&$1$&$1$&$240(4)$&$240(4)$&$288(3)$&$0.026$&$0.084$&$0.012$&$0.004$&$136$&$1.69$&$5.67$ \\ 
		$1\times10^9$&$1$&$\infty$&$1$&$1$&$240(4)$&$240(4)$&$288(3)$&$0.030$&$0.331$&$0.005$&$0.005$&$135$&$1.77$&$5.50$\\ 
	\end{tabular}
	\caption{Continue of table \ref{tab:case1}. }
	\label{tab:case2}
	\end{ruledtabular}
\end{table}

\end{document}